\documentclass[twocolumn,superscriptaddress,secnumarabic,amssymb,nobibnotes,aps,prx]{revtex4-2}

\usepackage{extarrows}
\usepackage{amsmath}    
\usepackage{graphicx}   
\usepackage{verbatim}   
\usepackage{color}      
\usepackage{subfigure}  
\usepackage[colorlinks=true, allcolors=blue]{hyperref}   
\usepackage{verbatim}   
\usepackage{longtable}
\usepackage{epsfig}
\usepackage{comment}
\usepackage{bm}
\usepackage{ulem}
\usepackage{dcolumn} 
\usepackage{textcomp}   
\usepackage{kotex}      
\usepackage[version=3]{mhchem}  
\usepackage{xcolor}
\usepackage{lineno}

\newcommand\redout{\bgroup\markoverwith
  {\textcolor{red}{\rule[0.5ex]{2pt}{0.8pt}}}\ULon}

\newcommand{\lno}{\ce{LaNiO2}}
\newcommand{\rno}{${\mathcal R}$NiO$_2$}
\newcommand{\lnof}{\ce{La3Ni2O5F}}

\begin{document}
\title{Anomalous behavior of the Ni$^{1+}$ moment  
in bi-infinite-layered La$_3$Ni$_2$O$_5$F
}
\author{Young-Joon Song}
\affiliation{Institut f\"ur Theoretische Physik, Goethe-Universit\"at Frankfurt, Max-von-Laue-Stra\ss e 1, 60438 Frankfurt am Main, Germany}
\affiliation{Asia Pacific Center for Theoretical Physics, Pohang, 37673, Korea}
\author{Warren E. Pickett}
\email{wepickett@ucdavis.edu}
\affiliation{Department of Physics and Astronomy, University of California, Davis, California 95616, USA }
\author{Kwan-Woo Lee}
\email{mckwan@korea.ac.kr}
\affiliation{Division of Semiconductor Physics, Korea University, Sejong 30019, Korea}
\affiliation{Department of Applied Physics, Graduate School, Korea University, Sejong 30019, Korea}
\date{\today}

\begin{abstract}
The discovery of superconductivity in hole-doped Ni$^{1+}$ systems with ``infinite layer" NiO$_2$ square-lattices analogous to the Cu$^{2+}$ CaCuO$_2$ cuprate has renewed conflicting pictures of the Cu$^{2+}$$-$Ni$^{1+}$ similarities and distinctions. Recent synthesis of formal Ni$^{1+}$ La$_3$Ni$_{2}$O$_{5}$F with two infinite NiO$_{2}$ layers per cell provides a novel member of this class.  We find that Ni$^{1+}$ in this material is unusually flexible. 
First principles density functional theory studies reveal a single partially occupied electron band derived from density in {\it three interstitial} layers that provides self-doping to a Ni$^{1.09+}$ charge state.
The blocking La(O/F)La layer provides isolation of the NiO$_2$ bilayer to strictly two-dimensional electronic and magnetic systems. 
Fixed spin moment calculations of magnetic tendencies reveal behavior unlike previous nickelates, including an essentially vanishing differential susceptibility up to a large magnetic field and a metastable ferromagnetic (FM) state. Antiferromagnetic (AFM) order is favored energetically as expected; surprisingly both FM and AFM states are obtained without correlation (Hubbard $U$) corrections. Doping away from half-filling by an interstitial density derived band and likely two-dimensional fluctuations account for the lack of observation of any magnetic transition. 
\end{abstract}

\maketitle
\clearpage

\section{Introduction}

After decades of effort to discover superconductivity (SC) in layered Ni$^{1+}$ square-lattice systems that are isostructural and isoelectronic to the parent compounds of high-$T_c$ cuprates \cite{levitz1983a,levitz1983b,Hayward1999,anisimov,LP2004,norman2020,wep2021}, 
in 2019 SC was discovered in hole-doped NdNiO$_2$ with critical temperature up to $T_c\approx15$ K \cite{hwang2019}. Since then, layered superconducting nickelates have attracted intense investigation. In hole-doped \rno~(${\mathcal R}=$La, Nd, Pr, Sm) systems  \cite{Osada2020,sw.zeng2022,s.zeng2022a,xiao2024} SC has been observed, with onset temperatures reaching $T_c\approx40$ K at ambient pressure for hole-doped SmNiO$_2$ \cite{chow2025}. SC has been reported also in nominally undoped members \cite{,parzyck2025,sahib2025}. Additionally, SC has been discovered in the Ruddlesden-Popper NiO$_2$ multilayered systems:
the quintuple-layer Nd$_6$Ni$_5$O$_{12}$ with $T_c\approx13$ K \cite{pan2022}, the tri-layer ${\mathcal R}_4$Ni$_3$O$_{10}$  members up to $T_c\approx30$ K under pressure \cite{y.zhu2024,e.zhang2025}, and in particular the  ${\mathcal R}_3$Ni$_2$O$_{7}$ members with substantially high $T_c$ of 80 K under pressure  \cite{n.wang2024,ey.ko2025,y.liu2025,b.hao2025}.
Despite furious experimental \cite{hepting2020,spinwave2021,nmr2021,nmr2021chin,Fowlie2022,d.li2020,s.zeng2020b,osada2020prm,k.lee2023,eren2025,z.dong2025,x.ding2023,krieger2022,x.ding2024,w.sun2025,c.li2025,rossi2022,tam2022,k.chen2024,khasanov2025} and theoretical \cite{LP2020a,LP2020b,botana2020,sawarzky2020,g.zhang2020,karp2020, k.held2020,arita2019,saka2020,katurkuri2020,vish2020,bandy2020, lecher2020,kang2021,choi2021,fp2021,w.ku2021,botana2024,j.zhan2025,abadi2025,ryee2025,bz.li2020,kh.kim2025} efforts to understand the SC, consensus of its origin and the unusual behavior of the Ni moment remains elusive. 

Behavior of the doped Ni$^{1+}$ ion lies at the base of this SC. However, a new feature is that \rno~materials have been established to harbour an interstitial density located at the open apical-O position. This band hole-dopes the system, greatly confusing its impact and indeed its origin. Such electron-as-anion materials have persisted in interest since the discovery of a crystal of double crown ether molecules, which in encapsulating a Cs ion expels its valence electron into the interstitial region \cite{dye1993,djsingh1993,wagner1994}. Several theoretical conjectures about the character and impact of this interstitial electride band (here called $E^*$ band) \cite{arita2019,nomura2022,adhikary,y.gu2020,j.you2026,sawatzky2023}, especially on the SC, have culminated in the determination \cite{w.sun2025} by angle-resolved photoemission spectroscopy (ARPES) that its circular Fermi surface (FS) agrees with calculations \cite{x.ding2024} and that the angular dependence of the $E^*$-band is circular \cite{c.li2025}. One evident impact is to provide some intrinsic electron self-doping of the cell, hence hole-doping of the Ni ion. 
Our studies demonstrate that the behavior of the Ni$^{1+}$ ion in \lnof~is distinct from any other nickelate, and that the character of the $E^*$ band has novel aspects due in part to the very effective blocking layer in this new compound.

\begin{figure*}[htb]
\includegraphics[width=2.0\columnwidth]{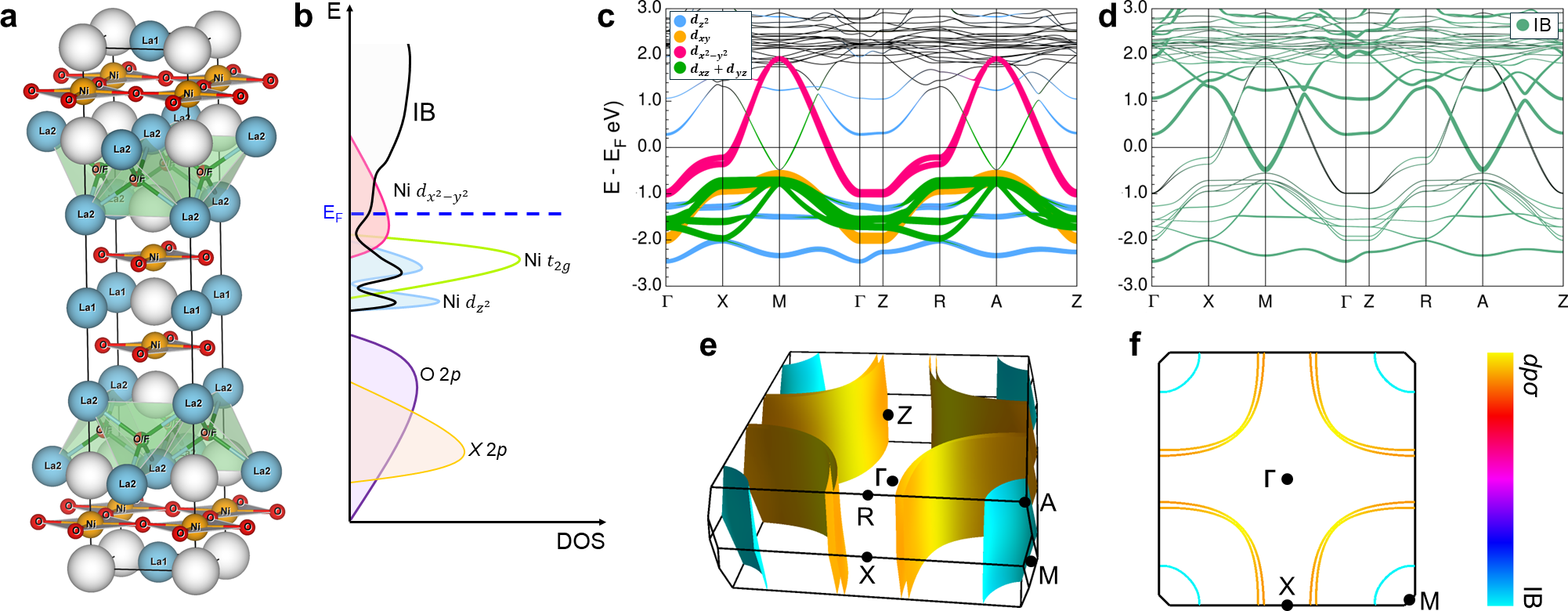}
\caption{{\bf a,} Tetragonal $I4/mmm$ structure of \lnof,
composed of bi-infinite-layer NiO$_2$ square lattices separated by the La1 ion. PbO-type [La2][$X$][La2] layers
provide a blocking layer that enforces extreme two-dimensionality (see text). Here, $X$ denotes sites equally occupied by O and F ions. 
White spheres are centered at the three apical sites and {\it do not} represent the shape of the interstitial density (see text and Fig.~\ref{fig2}).
{\bf b}, Sketch of the orbital-projected densities of states (DOSs) for Ni $3d$, O and $X$ $2p$, and interstitial $E^*$ band (IB) 
in the NM state at the GGA level.
{\bf c,}{\bf d,} GGA band structure  
in the range of $-3$ to 3 eV, with the fatband plots of ({\bf c}) Ni $3d$ and ({\bf d}) the $E^*$ band that dips below E$_F$ at the $M(A)$ point. The $E^*$ fatbands reflect the density throughout the interstitial region, {\it i.e.} not within any atomic sphere. The empty $4f$ orbitals of La ions are located in the 2--3 eV range. Symmetry points follow those of the primitive tetragonal Brillouin zone (BZ), as given in the corresponding FSs. 
{\bf e,}{\bf f,} 3D and 2D-projected FSs colored by orbital characters, showing the relative contributions of $dp\sigma$ and interstitial $E^*$ band.
These large surfaces show extreme 2D character, and the small cylinders circling $M$-$A$ arise from the $E^*$ band.
}
\label{fig1}
\end{figure*}

As pointed out in our previous works \cite{LP2004,LP2020a,LP2020b} 
and confirmed by others, the Ni$^{1+}$ ion behaves differently from the Cu$^{2+}$ ion, despite both being formally $3d^9$ configurations. Although substantial antiferromagnetic (AFM) interaction in \rno~was indicated by several experimental observations \cite{spinwave2021,nmr2021,nmr2021chin,Fowlie2022}, no AFM order has yet been observed, and has required correlation corrections in theory to describe Ni moments. Most indications are that the Ni exchange coupling is substantially smaller than in cuprates.

Recently Wenert, Smyth, and Hayward reported synthesis of the bi-infinite-layered Ni$^{1+}$ system of \lnof~\cite{hayward2026}, establishing the structure but leaving the magnetic behavior unclear, due to ferromagnetic (FM) Ni inclusions 
that complicate the susceptibility data. AFM order was suggested, but in the measured and extracted $M/H$ data there was no sign of any magnetic phase transition.
In isostructural insulating Ni$^{1.5+}$ La$_3$Ni$_2$O$_6$ \cite{poltavets2006}, similarities with superconducting Cu$^{2+}$/Cu$^{3+}$ cuprates were noted \cite{poltavets2009}.
No AFM phase transition was detected down to 4 K in the polycrystalline sample \cite{poltavets2009}, but a weak susceptibility anomaly in a single crystal was suggested as evidence for (charge or spin) density waves \cite{z.Liu2023}, potentially linked to theoretically proposed Ni$^{1+}$/Ni$^{2+}$ charge ordering \cite{pardo2011,botana2016}.

The focus of this paper is the unusual magnetic behavior shown by \lnof. In a sister paper \cite{Song2-2026} we have described and analyzed the unexpected electron band, and circular FS of, that is unparalleled in nickelates. While it is nearly impervious to the magnetic behavior we describe here, this band displays linear dispersion and lies in the vicinity of a Dirac point in partnership with Ni $3d$ bands, requiring a separate publication.

\section{Crystal Structure,\\ Calculational Methods}

From the crystal structure of Fig. \ref{fig1}a, the separation of the NiO$_2$ bilayers by the PbO-type [La2][$X_2$][La2] blocking layer is evident, in this way \lnof~is different from isostructural La$_3$Ni$_2$O$_6$ by replacement of half of the O ions in the blocking layer by F, and of course the extra hole. Also shown by the white spheres are the empty apical sites/cell, which attract density due to a favorable Madelung potential and to open space allowing decrease of the kinetic energy. The experimental lattice parameters $a=3.9839$ \AA~and $c=19.295$ \AA~\cite{hayward2026} lead to Ni-O distance $d_\parallel(=\frac{a}{2})$ of 1.992 \AA~ and out-of-plane Ni-Ni distance $d_\perp$ of 3.253 \AA. The ratio 
$\frac{d_\perp}{d_\parallel}\approx1.633$ can be compared with 1.744 in \lno~\cite{Hayward1999,LP2004}, so the NiO$_2$ layers are closer than in infinite-layer nickelates (ILNs) materials.

We have investigated the electronic properties and magnetic behavior of \lnof, first using the generalized gradient approximation (GGA) exchange-correlation (XC) functional \cite{gga} 
and secondly the GGA+U approach to include correlation effects, considering the effective $U_{eff}=U-J_H$ on Ni ions with the Hubbard repulsion $U$ and the Hund’s integral $J_H$.
In all calculations the $X$ $2p$ orbitals are fully occupied, lying in the $-6$ to $-3$ eV range below the Fermi energy E$_F$ or the gap, and the La $5d$ orbitals are unoccupied. Both all-electron full-potential codes {\sc wien2k} \cite{w2k} and {\sc fplo} \cite{fplo1} have been used in various calculations as noted.

\section{Nonmagnetic electronic structure}
\subsection{Band structure}
The nonmagnetic (NM) band structure is the reference for considering the spin-less or spin-fluctuation systems as \lnof~appears to be, and is often in agreement with ARPES of the FS for both nickelates and cuprates. Moreover, the self-doping that drives the hole band away from half filling and the extreme two-dimensional (2D) character, to be described below, is enough to account for the lack of magnetic ordering. While some aspects are similar to those of \lno, several others in \lnof~are qualitatively different.

In \lnof~the bands in the neighborhood of E$_F$ are 2D to any physically relevant measure: $k_z$ dispersion of 1 meV or less \cite{Song2-2026}, then perpendicular magnetic exchange across the blocking layer would be $J_{\perp}=4t_{\perp}^2/U$$\sim$ 1$\mu$eV for $U$=4 eV. In ILNs, with no blocking layer, $k_z$ dispersion of all bands is part of the full picture. The La$X_2$La blocking layer in \lnof, along with the body-centering structure, is highly effective in isolating the conducting part of the cell from neighboring bilayers. The Ni fatbands representation is shown in Fig. \ref{fig1}c; the two Ni $d_{x^2-y^2}-$O ${p_\sigma}$ bands (hereafter ``$dp\sigma$") are essentially degenerate (non-interacting) except when they approach the $X$ point. They cross $E_F$ in the usual near-half filling fashion.

The $dp\sigma$ pair of bands, including the splitting at the $X$ point, are reproduced in a 2$\times$2 tight binding model, 
with on-site energy 
$\varepsilon_o$=0.087 eV (relative to $E_F$=0) and hopping parameters 
$t_{100}$=$-0.363$ eV,
$t_{110}$=$0.094$ eV,
$t_{\perp}$=$0.039$ eV,
with the intra-bilayer hopping $t_{\perp}$ contributing to the splitting that is visible at $X$ and $R$ in Fig.~\ref{fig1}c,d. For comparison, for LaNiO$_2$ we had obtained (using the current sign conventions, see in the supplementary information (SI))
\cite{LP2004} for in-plane dispersion
the similar values 
$\varepsilon_o=0.093$ eV, 
$t_{100}$=$-0.381$ eV,
$t_{110}$=$0.081$ eV, 
$t_{\perp}$=$0.058$ eV.
The main differences are the lack of $k_z$ hopping and the larger $t_{\perp}$ in \lnof,~and the band filling. 

The splitting at $X$ and $R$ gives rise to the two van Hove singularities (vHs) around $-0.2$ eV and $-0.3$ eV in the DOS, shown in Fig. S3 (SI), that figure in the magnetic calculations.
The $dp\sigma$ FSs might be considered as nearly degenerate squares with greatly rounded corners or alternatively as approximately circular (see Fig. \ref{fig1}f). $k_z$ dispersion of both the $dp\sigma$ and $E^*$ bands is negligible (less than 1 meV). These details are pertinent, because the shapes, separations, and dimensionality of FSs are important to models of exotic superconducting order parameters. 
For future use and analysis, we find from comparing core energies, presented in Table S1 (SI), in {\sc wien2k} \cite{w2k} that the La2 ion, part of the blocking layer, has a lower site energy than La1 by 0.43 eV.

This $dp\sigma$ band is however not half-filled, hence less susceptible to Mott insulation and magnetic ordering, due to an interstitial electron density related band ($E^*$) of width 2.5 eV that dips 0.5 eV below E$_F$ along the $M$-$A$ line. This doping brings E$_F$ closer to the (two split by interlayer coupling) $dp\sigma$ vHs at the $X$ point. Both hole and electron FSs are displayed in Figs. \ref{fig1}e and \ref{fig1}f, showing that this $E^*$ cylinder FS contains 9\% of the zone area (0.18 electrons), giving an adjusted valence Ni$^{1.09+}$.
The FS radius around the $A$ point is similar to that of the LaNiO$_2$ ARPES data \cite{x.ding2024}. However, the $k_z$ dispersion of the electron band (their $\beta$ band \cite{x.ding2024}) cuts off that FS before halfway toward $M$. The resulting 3D electron FS, with incomplete data about its full boundary, makes experimental evaluation of the direct volume of the electron FS (the amount of self-doping) uncertain. Their value was obtained from the volume of the large $dp\sigma$ FS, for which experimental uncertainty of the radius and the $k_z$ dispersion leaves their value somewhat imprecise. The close agreement of their data with density functional theory (DFT) results allows the estimate that their $\beta$ FS contains $\sim$1/3 the volume of the $E^*$ FS of \lnof, however this latter doping must be divided between the two Ni atoms in the unit cell.  Thus our value of 0.09/Ni, from our analysis, indicates for \lnof~a somewhat larger self-doping per Ni than in LaNiO$_2$. The distribution of the electron density is however quite different, as we now describe.

\begin{figure}[tb]
\includegraphics[width=0.8\columnwidth]{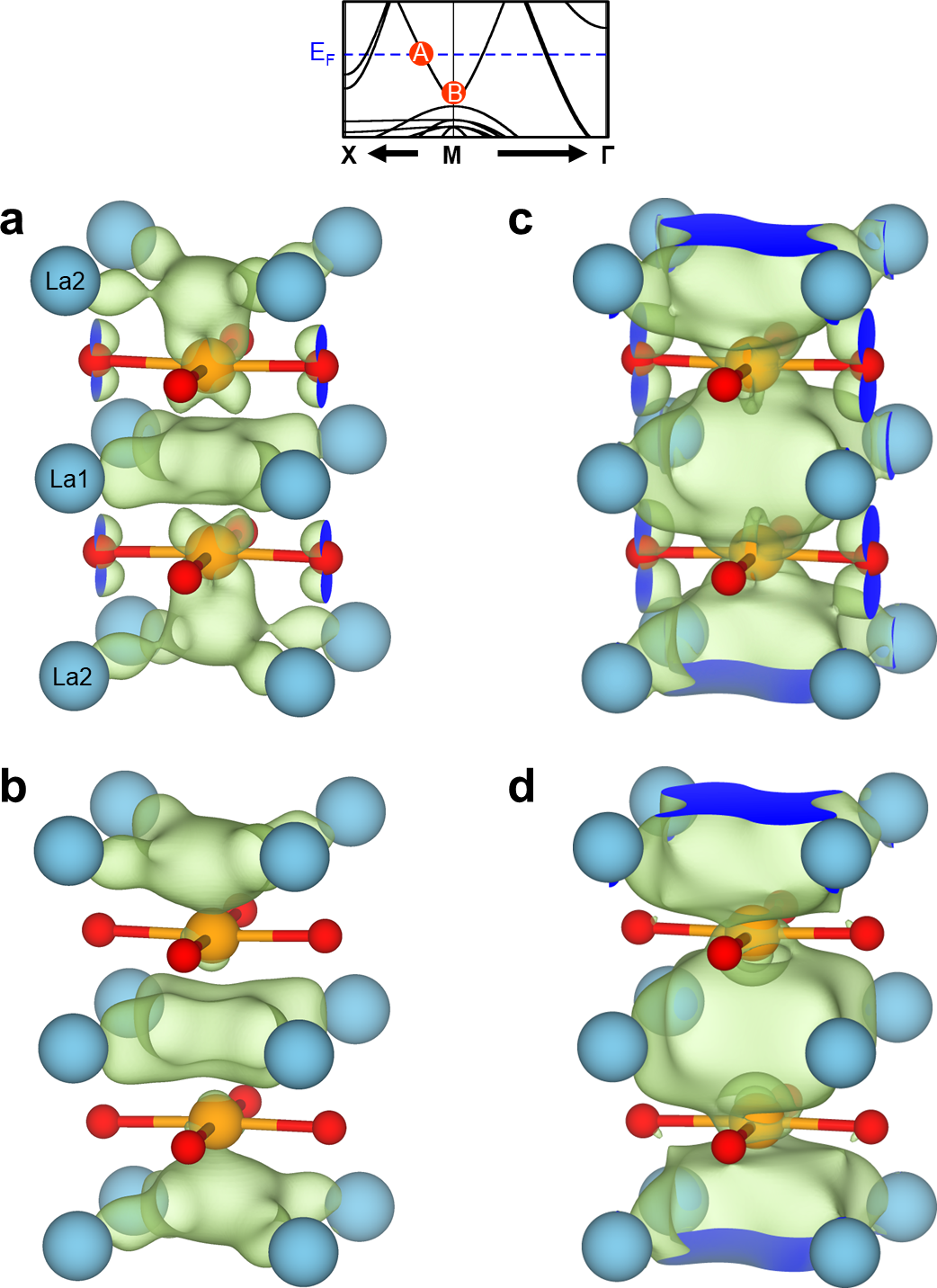}
\caption{{\bf Isosurface plots (lime-colored) of the $E^*$ band wavefunction $|\Psi({\bf r})|$ in the NM state.}
{\bf a,}{\bf b}, Using two different $\boldsymbol{k}$ points depicted at the top inset. {\bf c,}{\bf d,} At the same $\boldsymbol{k}$ points as ({\bf a}) and ({\bf b}), respectively, but with a 40\% lower isovalue, allowing an additional view of the $E^*$ shape. No detectable distinctions were found at the same $\boldsymbol{k}$ points as ({\bf a}) and ({\bf b}), but on the $\frac{\pi}{c}$ plane, i.e., near the A point. See the text for further discussion.
}
\label{fig2}
\end{figure}

\subsection{Interstitial band}
As mentioned, this $E^*$ band, with position and character discussed below and analyzed more fully in a separate paper \cite{Song2-2026}, is analogous to, but quite different from, the interstitial band seen in \rno~materials that have stimulated several studies \cite{arita2019,adhikary,y.gu2020,nomura2022,sawatzky2023,j.you2026}. In the \rno~materials where the interstitial density has been referred to an empty sphere (``s orbital'') the interstitial density is spherical or nearly so, whether plotted directly \cite{adhikary,y.gu2020,j.you2026} or via a maximally localized Wannier function based on an empty site orbital \cite{sawatzky2023}. However, direct plots of Bloch state densities \cite{arita2019,nomura2022} present anisotropic densities that are very different for near E$_F$ states at $\Gamma$ (La centered and $5d$ related) and midway between $R$ and $A$ (apical site centered). 

The real space picture in \lnof~is qualitatively distinct in the following ways (see Fig. \ref{fig2}): 
(i) the density is not confined to the empty apical O site, instead it is spread out over the interstitial region in each of the three La layers, yet it is a single band, see Fig.~\ref{fig1}d,
(ii) due to large $k_z$ dispersion (no blocking layer) the minimum of the electron band in ILNs is only around the $A$ corner of the zone, whereas in \lnof~the band minimum is flat along the entire $M$-$A$ line, 
(iii) the band is essentially linear from the bottom at $-0.5$ eV at the $M$ upward for 2 eV, with one implication being that it cannot be modeled with a hopping parameter or a Fourier expansion, 
(iv) because it has constant magnitude of velocity over this range, the effective mass tensor is purely off-diagonal, since the longitudinal effective mass is divergent,  
(v) it forms a cylindrical FS around the $M$-$A$ line, see Fig.~\ref{fig1}e. This $E^*$ band is effectively a positive energy Dirac band (upward in energy from an impending Dirac point \cite{Song2-2026}). 
Some of these features of the $E^*$ band will become important in the magnetic behavior discussed below.

Isosurfaces of the $E^*$ wavefunctions $|\Psi({\bf r})|$ at two $\boldsymbol{k}$ points are presented in Fig.~\ref{fig2}.
Each bilayer is quantum confined by the blocking layers,
allowing $k$-space attention to be confined to the basal plane. 
The shapes of the $E^*$ density have a broad maximum density at each apical site, with four fat ``windmill'' arms extending in [110] directions toward the La$^{3+}$ ions in each layer. 
For Figs. \ref{fig2}c and \ref{fig2}d the $\boldsymbol{k}$ points and wavefunctions are the same as in Figs. \ref{fig2}a and \ref{fig2}b, but using a smaller isovalue corresponding to a lower density. It then becomes evident that the densities of the left panel, corresponding to three fat ``windmills'', are connected through Ni $3d$ orbitals as  described in Ref. \onlinecite{Song2-2026}.
With this position and shape of interstitial density it is natural that there is some mixing with La $d_{xy}$ (same layer) and Ni $d_{z^2}$ orbitals (oriented toward the apical site). However, these orbitals are 3 eV above, respectively 1.5 eV below E$_F$, and being so distant in energy the mixing around E$_F$ is secondary.

\begin{figure}[tb]
\includegraphics[{width=0.99\columnwidth}]{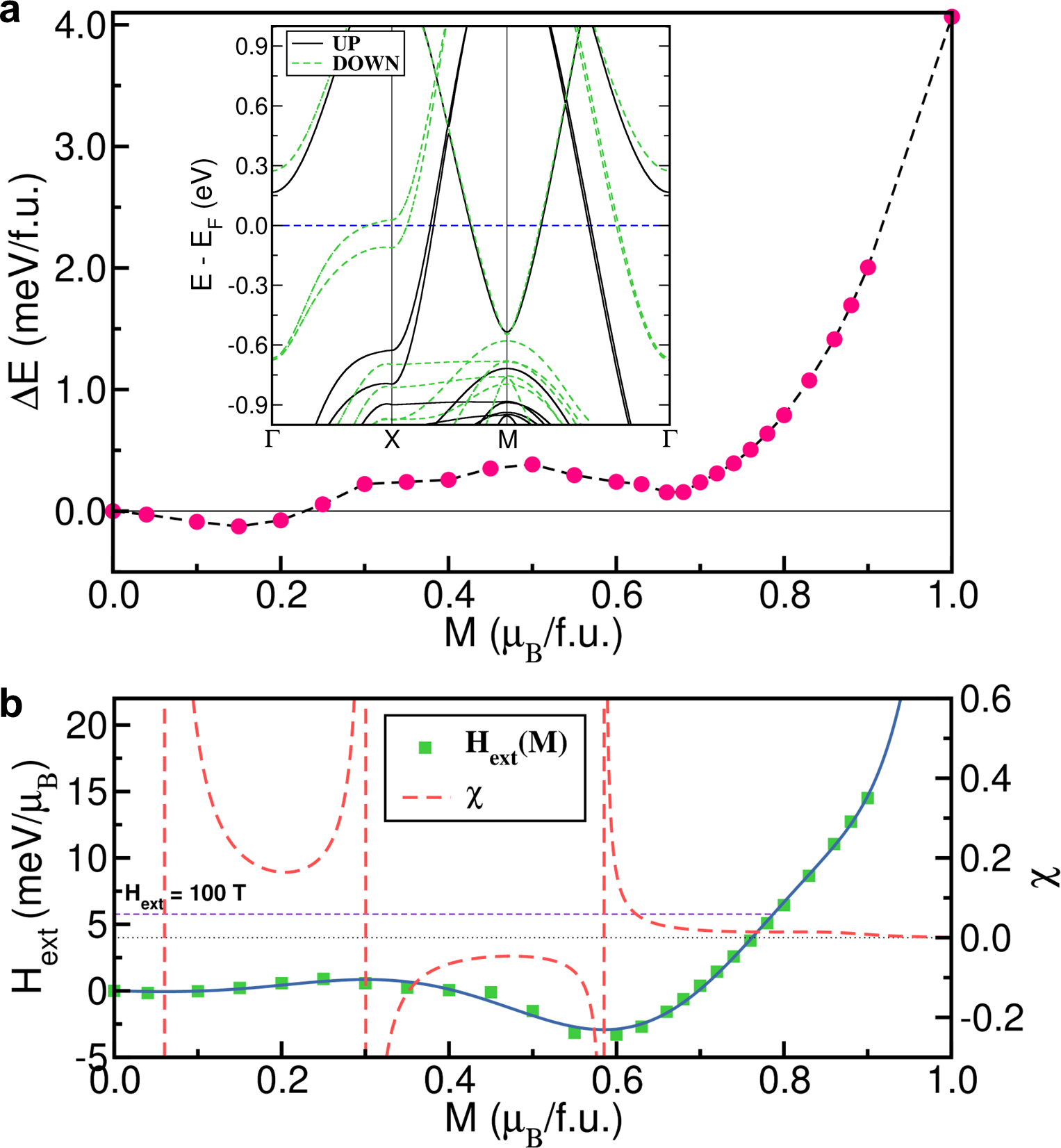}
\caption{{\bf Variations obtained from the constrained moment $M$ in the FSM calculations using {\sc fplo}.}
{\bf a}, Energy variation $\Delta E(M)$. 
Only above $M=0.8\mu_B$ does the magnetic energy begin to increase rapidly. The calculated values at $M$=0.2 and below, $\Delta E$=0.1 meV or less, indicate the limit of precision of this calculation.
{\bf Inset:} GGA FM band structure, showing the 0.6 eV exchange splitting at $X$, where the constrained FM moment is at the small peak in $\Delta E(M)$ at 0.5$\mu_B$, 
and a Lifshitz transition is occurring in the minority FS. Note the lack of any visible exchange splitting for the $E^*$ band (uppermost at $M$), and the lack of mixing with the exchange-split $dp\sigma$ bands along both directions around +0.5 eV.
{\bf b,} Calculated external magnetic field $H_{\rm ext}(M)$ (in meV/$\mu_B$), 
which requires less stringent numerical precision than does $\Delta E(M)$. 
The unphysical negative value of $\chi$ for $M$=0.3--0.6 indicates magnetic instability in that region, an inaccessible region. The right side indicates the differential susceptibility $\chi=(dH_{\rm ext}/{dM})^{-1}$, obtained from the fitting of $H_{\rm ext}(M)$ (blue-line).
}
\label{fig3}
\end{figure}

\section{Magnetic behavior}
\subsection{Ferromagnetic order}
\subsubsection{Ferromagnetism within GGA}
Although magnetic order has not been reported in Ni$^{1+}$ nickelates, GGA calculations have sometimes obtained AFM ordered states \cite{LP2004}, 
allowing estimates of the exchange coupling strength, whereas correlated calculations (DFT+U or DFT+manybody) are required to obtain Mott insulating splitting of the $dp\sigma$ bands.
Due to the 2D nature of \lnof~and its self-doping away from half-filling, both of which will oppose magnetic order, we do not expect magnetic order to be observed in \lnof~and none was seen in the initial report \cite{hayward2026}.  
However, to quantify magnetic tendencies we have first considered possible FM ordering. Defying expectations, 
unusual FM behavior emerges 
that distinguishes the magnetic behaviors of the $dp\sigma$ and $E^*$ bands.

On general principles based on the in-plane 
AFM superexchange in nickelates and cuprates, FM alignment has been out of the question. Unexpectedly, in a self-consistent calculation starting from large Ni moments, a persistent self-consistent FM state is obtained in which the $dp\sigma$ bands are split by 0.7 eV, associated with a moment of 0.68 $\mu_B$/f.u. (see the Inset of Fig. \ref{fig3}a); 0.34$\mu_B$ is small for a Ni moment, more indicative of an itinerant moment than a local moment. The energy is above that for $M$=0, though by only 0.3 meV, indicating a  metastable state. This result is in sharp contrast to LaNiO$_2$ \cite{LP2004} and CaCuO$_2$ \cite{Singh-Pickett} (and other cuprates) in which a magnetic state emerges only when AFM alignment is considered \cite{LP2004}, requiring a Hubbard $U$ enhancement. This FM state prompted us to study FM tendencies more closely.

\subsubsection{Fixed spin moment studies} 
Peculiar behavior of Ni magnetism in \lnof~is illuminated by fixed spin moment (FSM) calculations \cite{fsm1,fsm2}. The Hamiltonian is appended by an external magnetic field ($H_{ext}$) term $-\mu_B\int m(\vec r)d^3r H_{ext}$, $m(\vec r)=n_{\uparrow}(\vec r)-n_{\downarrow}(\vec r)$, $n(\vec r)=n_{\uparrow}(\vec r)+n_{\downarrow}(\vec r)$ with the constraints 
\begin{equation}
\mu_B\int m(\vec r;N)d^3r=M;~~\int n(\vec r;M)d^3r=N,
\end{equation}
in terms of the spin-resolved densities $n_\sigma(\vec r)$. $H$ is varied until the constraint on $M$ is satisfied. The external field impacts regions greatest where the polarization is largest, hence it is not a rigid shifting of bands by $\pm\mu_BM$ as in the Stoner picture. Results from {\sc fplo} \cite{fplo1} using the GGA functional \cite{gga} are shown in Fig. \ref{fig3}a. Up to $M$=0.7 $\mu_B$ there is extremely little energy increase, less than 0.3 meV/f.u above $M$=0. The Ni FM moments (implying also long wavelength fluctuations) cost little or negligible energy, indicating that the system is almost upon a quantum critical point (magnetic ordering), seemingly unrelated to a large value of DOS N($E_F$) at $E_F$, {\it i.e.} the Stoner picture. 

To calculate the usual magnetic energy expansion $\Delta E(M)=a_2M^2 + a_4M^4 + \cdots$ is found to be unusually challenging for \lnof; the expansion only holds for ``small $M$'', however the $a_2$ coefficient we find is exceedingly small, making it subject to numerical imprecision, choice of XC functional \cite{Larson2004}, and if changes are exceptionally small, the manner of implementation such as basis set (in)completeness. Finding $a_2$ to be below numerical precision puts the system at a quantum critical point (QCP) with respect to FM order. Convergence of the calculation and limitation of the expansion is further complicated by a minority vHs passing through $E_F$ in the region of interest, see the Inset of Fig.~\ref{fig3}a. Meanwhile, as the Ni moment is increased to $\sim$0.7$\mu_B$ by (small) $H_{ext}$, the $E^*$ FS is only slightly polarized by the small splitting (invisible in plots) from Pauli paramagnetism $\mu_B N_{E^*}(E_F)$, and unenhanced due to the absence of a Hund's coupling (Stoner $I$) in the interstitial density. In the following discussion we make use of the resulting external field $H_{ext}(M)$ from FSM calculation that is necessary to produce magnetization $M$, as it is less susceptible to numerical imprecision \cite{Larson2004}.

The weak maximum in $E(M)$ at $M$=0.5$\mu_B$ is due to a minority vHs passing through $E_F$, indicating a Lifshitz transition. This topological change in the FS will involve small to near vanishing velocities in the neighborhood of the $X$ points. There is a minimum (metastable state) at $M$=0.68$\mu_B$ corresponding to the above mentioned FM solution. Both the extremely small energy change $\Delta E(M)$ and the finding of a metastable FM state are unknown occurrences in nickelates.

An alternative to $\Delta E(M)$ is to study 
$H_{ext}(M)$, formally equal to $d\Delta E(M)/dM$ but known to be less noisy \cite{Larson2004}. The FSM behavior of the external field $H_{ext}(M)$ is shown in Fig.~\ref{fig3}b, with calculated values fit to an eleventh order in $M$ to systematize the structure in $H_{ext}(M)$.
Between $M$=0.3$\mu_B$ and 0.7$\mu_B$, the decrease in $E(M)$ indicates an unphysical regime in which the field $H_{ext}(M)$ decreases as the moment $M$ increases, a phenomenon that is not uncommon in FSM calculations \cite{fsm1,fsm2} of weakly or nearly FM materials with unstable regions. 

The differential susceptibility $\chi(M)=(dH_{ext}/dM)^{-1}$,
also shown in Fig.~\ref{fig3}b, encounters poles at $M$=$0.3\mu_B$ and $M$=0.7$\mu_B$ -- inflection points of $E(M)$ -- between which the unphysical negative $\chi$ indicates an inaccessible region. (The pole at 0.06$\mu_B$ is a numerical inaccuracy where $H_{ext}(M)$ is flat to numerical precision). $H_{ext}(M=0.68\mu_B)$ is zero at the minimum of $\Delta E(M)$, the metastable FM state discussed in the previous subsection. At such a point $M$ can increase without any first order change in energy -- $\chi$ diverges. From the AFM energies given in the next subsection, this metastable FM state is unlikely to be attainable, by somehow becoming stuck in the minimum during  synthesis. However, application and then release of a laboratory field $H_{ext}$$\sim$10T might allow the sample to settle into this local minimum (as the DFT GGA calculation does). In addition, in the paramagnetic state at $H_{ext}$=0 \lnof~surely will have (disordered) Ni local moments with a large susceptibility, which is not a part of the current analysis, but see Sec. \ref{sec:flucs} below.

The mysterious behavior of $E(M)$ and $H_{ext}(M)$ in \lnof~indicates yet unseen magnetic behavior. The FSM energy change involves three contributions
$\Delta E(M) = E_K(M)+E_P(M)+E_{XC}(M)$: respectively, the true kinetic energy $E_K$ cost [from $-(\hbar^2/2m)\nabla^2$, not as often envisioned as the (Stoner) band energy which double counts potential terms]; the conventional ``exchange'' polarization energy $E_{XC}$; and the little recognized potential term $E_P(M)$, the change in the Hartree+external (nuclear) energies due to variation of the {\it total density} $n(\vec r;M)$ with $M$.  

This latter term $E_P[n(M)]$ is usually passed over (at least in understanding) as the density is expected to show insignificant variation with $M$. However, the $dp\sigma$ band density is evolving in its fractions of $d_{x^2-y^2}$ and $p_{\sigma}$ orbital character across the band. Plots of these density changes, via wavefunction character, are provided in Fig.~S4 of SI. This change in density  across $E_F$, which is the region that is involved in $\Delta E(M)$, gives a gain in magnetization energy comparable to -- actually twice as large as -- the $E_{XC}$ term. Together they conspire to cancel the $E_K$ kinetic term (quite precisely), which would normally be expected to dominate for nickelate (and cuprate) $dp\sigma$ bands forced toward FM, when their AFM exchange term normally dominates. The separation into $E_K(M)+E_P(M)+E_{XC}$ components is plotted in Fig. S5 of SI, with further description. 

For comparison, for the typical Ni Stoner constant $I$$\approx$0.5 eV and this moment of $M$=0.34$\mu_B$/Ni, the Stoner exchange energy/f.u. is twice (two Ni atoms) $-\frac{1}{4}$$I$$M^2$ = $-28$ meV, to be compensated by the band energy cost. The DFT FSM energies are, at $M$=0.68$\mu_B$/f.u.:
magnetic energy gain of $-419.7$ meV/f.u. (1/3 from $E_{XC})$, 
kinetic energy cost of 419.5/f.u.,
net energy of 0.1 meV/Ni as shown in Fig. \ref{fig3}a, and indicating just how delicate the energy balance is.
This XC magnetic energy term is a factor of five larger than the Stoner model estimate, but does not help to explain such precise cancellation of energies. 
Back to the FM energy of \lnof, the FSM magnetization energy only begins to increase sharply beyond $M$=0.8$\mu_B$ ($M_{\rm Ni}$=0.4$\mu_B$) -- the kinetic energy dominates. Results in the next subsection indicate that the FM state is unstable with respect to AFM states, which have a factor of two larger Ni moments.

\begin{figure}[tb]
\includegraphics[width=0.9\columnwidth]{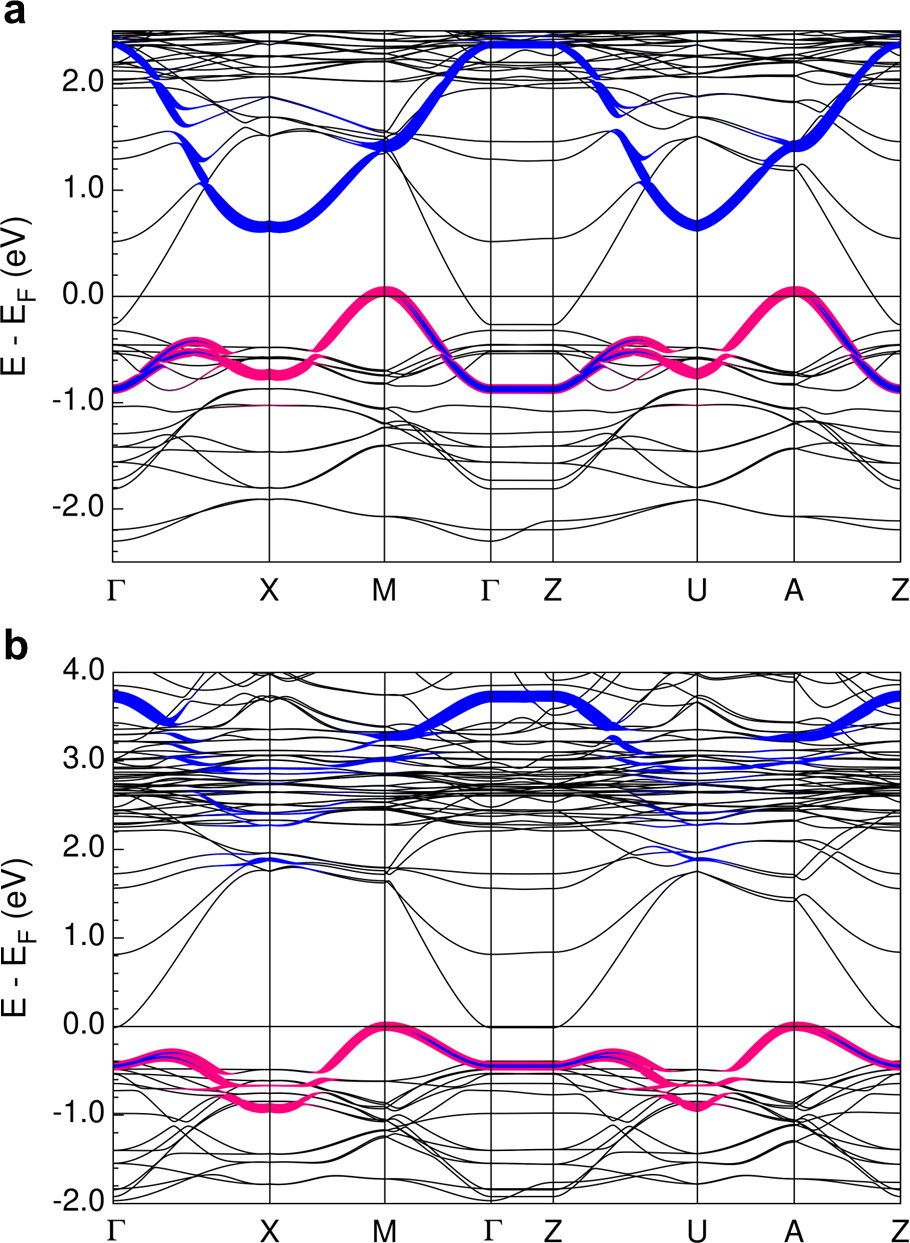}
\caption{G-AFM fatband plots of ({\bf a}) GGA and ({\bf b}) GGA+U ($U_{eff}$=3 eV) for \lnof, in the range covering the spin-split $dp\sigma$ bands (red- and blue-colored for spin up and down, respectively). The splitting of the GGA+U is $\sim$3.5--4 eV, increased from the GGA value of 2 eV. At the GGA+U, 
the bottom of the $E^*$ band ($\Gamma$-$Z$, folded back by AFM order), now empty, is degenerate with the top of the majority $dp\sigma$ band, now a zero-gap semimetal band structure. Larger $U$ opens a gap, but slowly. 
Note the extreme flatness of the $E^*$ band along $\Gamma$-$Z$ -- no dispersion.
}
\label{fig4}
\end{figure}

\subsection{Antiferromagnetic order}
 For single layer CaCuO$_2$ with the same $d^9$ configuration, extensive efforts to reproduce the observed AFM ordering using GGA were unsuccessful \cite{Singh-Pickett}.
However, for \lnof~AFM states were readily obtained within GGA for both the aligned (C-AFM) or antialigned (G-AFM) AFM NiO$_2$ layers.
For energies from {\sc wien2k}, G-AFM is lower energy than NM, FM, and C-AFM by 155, 159, 3.4 meV/f.u., respectively.  
The Ni moment in the atomic spheres for both AFM orders is 0.68 $\mu_B$, doubled relative to the FM moment discussed above and indicative of a spin-half local moment reduced by hybridization with O $p_{\sigma}$ orbitals. Since AFM order is energetically favored in all calculations (which as commonly done at this level of neglect of fluctuation effects), it requires the primary attention.  

For the two AFM alignments, the self-doping aspect, {\it i.e.} the relative position and filling of the $E^*$ band, is decreased, but only slightly. For G-AFM at the GGA level, the $dp\sigma$ orbitals are already exchange split by $\sim$2 eV; see the fatband plots in Fig. \ref{fig4}a. 
This large and unexpected exchange splitting is more characteristic of a Mott splitting than an itinerant AFM splitting. In fact this result would closely emulate a Mott insulating state except for the $E^*$ band sweeping through the exchange (``DFT Mott'') gap, providing self-doping and electron carriers. The top valence bands have majority $dp\sigma$ character and slightly cross E$_F$ giving small hole cylinders (see Fig. S11 of SI). The electron tube around the $\Gamma-Z$ line is the $E^*$ FS, and self-doping is only slightly smaller than for NM and FM levels. Self-doping is only affected by correlation effects, as discussed next.

\section{Correlation effects: DFT+U}
Unlike other nickelates and cuprates, including the Hubbard $U$ on FM Ni leads only to additional Mott-like splitting of the (already in GGA) magnetic band, and an increase in a Ni moment to 0.85$\mu_B$, more representative of a $S=\frac{1}{2}$ ion\ in an oxide.  Inclusion of $U$ reduces the energy difference between FM and AFM states: at $U_{eff}=3$ eV, the G-AFM state is still energetically favored, by 61 (2.2) meV/f.u. over FM (C-AFM) alignment [compare for GGA: 159 (3.4) meV/f.u.]. The FM-AFM energy difference decreases with further increase in $U$, indicative of a more localized moment.  For the effect of $U$: for $U_{eff}=4$ eV, the G-AFM state is still energetically favored, by 28 (2.4) meV/f.u. for the FM (C-AFM) states. 

For $U_{eff}$=3 eV, energy differences and applying the usual Heisenberg form of energy $-\frac{1}{2}\sum J_{ij}S_iS_j$, the exchange parameters can be estimated: in-plane $J_{\parallel}\approx$118 meV (1370 K), intra-bilayer $J_{\perp}\approx$4.5 meV, 4\% of $J_{\parallel}$ indicating weakly coupled AFM layers. Estimations for the ILNs vary but lie in a somewhat lower range. Although the 3D character should encourage ordering in ILNs no magnetic transition has been reported.

While the self-doping level (Ni valence) is unaffected  by both FM and AFM alignments in GGA, applying the Hubbard $U$ does have an effect. Due to the relative position of the $E^*$ band bottom (folded back to $\Gamma$ by AFM order) and the majority $dp\sigma$ band, $E^*$ band doping decreases with increasing $U$ for G-AFM. This change is likely best considered as due to lowering of the majority Ni band by application of $U$, rather than any upward shift of the $E^*$ band; polarization of the $E^*$ band is never noticeable. 

Only at $U_{eff}$=3 eV is a gap on the verge of opening, as displayed in Fig. \ref{fig4}b: the lower conduction band is then solely the interstitial $E^*$ band -- not a Mott insulator, rather a ``multiband Mott semimetal'' in a sense, however, Ni charge has changed to Ni$^{1+}$ by the emptying of the $E^*$ band. The split $3d$ band leaves a band of unpolarized, quasi-free electron like $E^*$ carriers that would be populated by further electron doping, {\it e.g.} increased F content, in the background of a Mott insulating Ni$^{1+}$ system. The relative band shift (from GGA) is changed by roughly half of $U_{eff}$ in the range we have considered. To emphasize: below $U_{eff}$=3 eV \lnof~is an unconventional $dp\sigma$ -- $E^*$ band semimetal, above this value the gap is between the $dp\sigma$ holes and unpolarized $E^*$ electrons rather than a Mott gap, and \lnof~becomes undoped -- a Ni$^{1+}$ ion again.

The $e_g$ splitting of G-AFM is different from that of isovalent \rno: NdNiO$_2$ shows at $E_F$, lying dispersionless, the Ni $d_{z^2}$ band along the $k_z=\frac{\pi}{c}$ zone face \cite{LP2020b}, while $d_{z^2}$ lies 1 eV lower in \lnof~and is inactive in magnetism but mixes strongly with the interstitial density \cite{Song2-2026}. This behavior reflects an increased  $e_g$ splitting in \lnof~and affects coupling (or lack thereof) along the $\hat{c}$ direction.  
Recall that $d_{z^2}$ had been involved at $E_F$ in \lno~\cite{LP2004}.
At different formal valence, La$_3$Ni$_2$O$_6$ has one more hole (thus Ni$^{1.5+}$), for which no SC nor any magnetic order down to 4 K has been reported. The additional hole per f.u. suggests that the $d_{z^2}$ orbital becomes more involved. Indeed, GGA+U calculations has indicated for La$_3$Ni$_2$O$_6$ a fully polarized $dp\sigma$ orbital and a Mott-split $d_{z^2}$ orbital \cite{pardo2011} (samples are insulating).
The extra hole provides a far different picture, reflecting strong doping dependence of the bilayer NiO$_2$ system.

\section{Spin fluctuation effects}
\label{sec:flucs}
Larson, Mazin, and Singh have translated Moriya's theory of quantum critical magnetic fluctuations near a quantum critical point to a first principles treatment of elemental spin-fluctuation enhanced  Pd \cite{Larson2004}. The corrections involve the first and second derivatives of the band energy $E_k$ at $E_F$, the first giving the FS velocity $dE_k/d{\vec k}={\vec v}_k$ and the second giving the inverse mass tensor $\mu^{\alpha,\beta}=d^2E_k/dk_{\alpha}dk_{\beta}$, as well as first and second energy derivatives of $N(E)$ at $E_F$, all requiring unusually fine sampling on and around the FS \cite{t.jeong}. 
This treatment can be applied to \lnof, which brings some simplifying aspects. 

Their treatment can be applied to \lnof, and a small approximation simplifies the result. 
Let us approximate the $dp\sigma$ NM FS as circular and isotropic, hence a constant velocity $v=|\vec v_k|$ over the FS, corrections should be second order in the anisotropy. This description is excellent for the $E^*$ band, however that band seems not to enter the magnetic behavior and we will not consider it. The expressions for determining the lowest order $q^2$ contribution to the susceptibility $\chi(q)$ depend on the derivatives mentioned above. For 2D isotropic dispersion $E_k=v|k|$ around the band center, $\vec v_k=v\hat{k}$, and $\mu^{\alpha,\beta}_k$ is a symmetric tensor with diagonal elements $\mu_k^{\alpha,\alpha}$=0 -- a divergent longitudinal effective mass -- and conventional off-diagonal matrix elements that do not enter lowest order spin fluctuation effects. With linear dispersion $N(E)=E/2\pi v^2$ per spin per unit volume, its E-derivative is constant and  its second derivative vanishes. Likewise, $\mu_{\alpha,\alpha}$ vanishes, and the coefficients of the $q^2$ terms in $\chi(q)$ -- the lowest order corrections to the band theory value  of the susceptibility from zero point fluctuations -- vanish. This result confronts the general expectation that fluctuations should be larger in 2D than in 3D, but requires the linear dispersion.  

There will be a contribution from the strict non-circularity of the $dp\sigma$ band, to second order in the non-circularity. Still, this conventional spin fluctuation renormalization of the magnetic response promises to be small. Broader considerations will be the extreme 2D character of the spin system, plus the spinwave scattering from the itinerant $E^*$ carriers, probably small in \lnof~ but perhaps important given that small effects may be crucial in a system very near a quantum critical point.

\section{Discussion and summary}
The behavior of the ``Ni$^{1+}$" ion in \lnof~is fundamentally different from other monovalent Ni oxides, and of course different from isostructural Cu$^{2+}$ in layered cuprates. The effect of the $E^*$ band seems small except when correlation effects for AFM order decreases its self-doping.  Within a first principles description, distinctive features that emerge are several: (i) a uniquely linear, partially occupied electron $E^*$ band arising from interstitial density (see Ref. \onlinecite{Song2-2026}) crosses the Fermi level giving self-doping, (ii) strong magnetic tendency for FM Ni polarization indicating close proximity to a quantum critical point, (iii) a metastable FM state with Ni moments of 0.34 $\mu_B$, (iv) energetically preferred AFM states. These latter are of the usual sort, except for the notable fact that they arise even without correlation effects (Hubbard $U$).

Energetics (neglecting fluctuations) support AFM order in-plane, intra-bilayer coupling is minor but favors antialignment of neighboring spins along $\hat c$ in the NiO$_2$ bilayer. Isolation of the electronic and magnetic states would normally suggest that 2D fluctuations will impair AFM order, viz. the Mermin-Wagner theorem \cite{mermin}. 
A relevant model comparison to \lnof~is the coupled square bilayer 2D Heisenberg system \cite{millis,l.yin1998,eshrig}. Like the single layer system with related fluctuations it does not order, but differs in detail by having a magnetic correlation length that diverges more strongly than in the single layer case \cite{hayden2004,stock}. 

Expressions from the Moriya theory of spin fluctuations near a quantum critical point developed by Larson {\it et al.} \cite{Larson2004} lead to the possibility that the linearity and isotropy of the $dp\sigma$ band near half-filling may degrade fluctuation effects greatly. The interplay between the $dp\sigma$ and $E^*$ bands, and implications of the strong two dimensionality, will benefit from further study.

Related comparisons may shed some light.
A single layer Ni$^{1+}$ oxyfluoride La$_2$NiO$_3$F was synthesized by Wissel {\it et al.} \cite{wissel2020} by topochemical removal of F from the AFM insulator La$_2$NiO$_3$F$_2$, and involving a structural reduction that can be described as \lnof~with missing La1 and one NiO$_2$ layer. Using first principles methods similar to ours, Bernardini {\it et al.} confirmed that the same blocking layer resulted in a 2D electronic structure \cite{bernardini2021}. A band that resembles our E$^*$ band barely touches E$_F$, and Wannierization reproduced that band by including additional La $d$ and O $p$ orbitals.  Their conclusion was that this single layer Ni$^{1+}$ ion behaves similarly to that of ILNs.

A well studied case is La$_2$CuO$_4$, with nearest neighbor $J_\parallel$$\sim$130 meV (and other intralayer couplings) \cite{coldea}. However the interlayer coupling along the $\hat c$ direction, facilitated by apical O, becomes crucial, enabling order with a resulting high Ne\'el temperature $T_N\approx$ 375 K. The explanation is that dynamic AFM regions tend to align along $\hat c$, re-enforcing even a small interlayer coupling to promote long range order. The bilayer AFM insulator YBa$_2$Cu$_3$O$_{6+\delta}$, with a blocking layer of (BaO)Cu(BaO) has been studied, again there is sufficient coupling through the blocking layer to enable AFM order in strongly underdoped (small $\delta$) samples \cite{kopp1996,hayden1996,millis}. Within the CuO$_2$ bilayer $J_{\parallel}$$\sim$120--150 meV and intrabilayer coupling $J_{\perp}/J_{\parallel}$$\sim$0.12 \cite{hayden1996,millis}, three times larger than for \lnof.

\vskip 6mm
{\large\bf Methods.}

 In the tetragonal structure (space group: $I4/mmm$, No. 139), La2 and Ni ions are located at $4e$ sites $(00z)$ with $z=$0.3212 and 0.0843, respectively.
 The other ions occupy the La1 $2b$ $(00\frac{1}{2})$, O $8g$ $(0\frac{1}{2}z)$, and $X$ $4d$ $(0\frac{1}{2}\frac{1}{4})$ sites.
In these calculations, 
the $X$ sites with 50\% O, 50\% F occupancy were treated within the virtual crystal approximation.
This is reasonable because the two atoms are adjacent in the periodic table and are located away from the NiO$_2$ layers.
For the AFM states
a $\sqrt{2}\times\sqrt{2}\times 1$ supercell was used.

Our calculations were performed with the accurate all-electron full-potential code {\sc wien2k} \cite{w2k}.
Some of the results were confirmed by another all-electron full-potential code {\sc fplo} \cite{fplo1}.  
Correlation effects were considered with the effective $U_{eff}=U-J_H=$3-5 eV on Ni ions, which is the proper range for the nickelates as observed previously \cite{LP2020a,LP2020b}.
The main features remain consistent in this range, although ionic band positions depend on the value of $U_{eff}$.
For the GGA+U calculations, two popular double-counting schemes of so-called ``around-mean field" (AMF) \cite{amf1993} and ``self-interaction correction" SIC \cite{sic1981,sic1994} were compared, resulting in similar trends likely because the $d^9$ formal valence does not leave flexibility in treating configurations. Unless stated otherwise, we addressed the results obtained from the AMF scheme.  
The basis in {\sc wien2k} was determined by $RK_{max}$=7.0 and atomic radii of 2.11 (Ni), O (1.65), $X$ (2.16) and La (2.39), in units of bohr radius. In {\sc fplo}, the default options were used. For the fixed spin moment calculations \cite{fsm1}, the BZ in the (nearly) itinerant system was carefully treated with a dense $k-$mesh of $30\times30\times30$. 

\vskip 4mm
{\large\bf Acknowledgments.}

We are indebted to K. Koepernik for detailed description and clarification of the fixed spin moment calculations in {\sc fplo}. Conversations with, and guide to the literature from, R. R. P. Singh are gratefully acknowledged.
KWL was supported by National Research Foundation of Korea (NRF) Grant (RS-2024-00392493). 
YJS was supported by the Deutsche Forschungsgemeinschaft (DFG, German Research Foundation) -- CRC 1487, ``Iron, upgraded!'' -- project number 443703006.

\bibliography{main}

\end{document}